\PassOptionsToPackage{svgnames}{xcolor}
\documentclass[twocolumn,times]{aastex62}
\usepackage{graphicx}
\usepackage[flushleft]{threeparttable}
\usepackage{blindtext}
\usepackage{amsmath}
\usepackage{mathtools}
\usepackage{multirow}
\usepackage{comment}
\usepackage{booktabs}
\turnoffeditone

\newcommand{\beq}{\begin{equation}}
\newcommand{\eeq}{\end{equation}}
\newcommand{\bea}{\begin{eqnarray}}
\newcommand{\eea}{\end{eqnarray}}
\newcommand{\Lya}{\ifmmode{\mathrm{Ly}\alpha}\else Ly$\alpha$\xspace\fi}

\begin{document}
\shortauthors{Park et al.}
\def\nar{New Astron.}
\def\na{New Astron.}
\title{\large \textbf{Analytic Model for Scattered Ly$\boldsymbol{\alpha}$ Emission in the Post-reionization Intergalactic Medium}}

\correspondingauthor{Hyunbae Park; Hyunmi Song}
\email{hcosmosb@gmail.com}

\author[0000-0002-7464-7857]{Hyunbae Park}
\affil{Center for Computational Sciences, University of Tsukuba, 1-1-1 Tennodai, Tsukuba, Ibaraki 305-8577, Japan}
\affil{Computational Cosmology Center, Lawrence Berkeley National Laboratory, CA 94720, USA}
\affil{Kavli Institute for the Physics and Mathematics of the Universe (Kavli IPMU, WPI), University of Tokyo, Chiba 277-8582, Japan}

\author[0000-0002-4362-4070]{Hyunmi Song}
\affil{Department of Astronomy and Space Science, Chungnam National University, Daejeon 34134, Republic of Korea}

\author[0000-0002-0885-8090]{Chris Byrohl}
\affil{Institut für Theoretische Astrophysik, Universität Heidelberg, ZAH, Albert-Ueberle-Str. 2, 69120 Heidelberg, Germany}
\affil{Kavli Institute for the Physics and Mathematics of the Universe (Kavli IPMU, WPI), University of Tokyo, Chiba 277-8582, Japan}

\author[0000-0002-2838-9033]{Aaron Smith}
\affil{Department of Physics, The University of Texas at Dallas, Richardson, TX 75080, USA}

\author[0000-0002-7464-7857]{Hidenobu Yajima}
\affil{Center for Computational Sciences, University of Tsukuba, 1-1-1 Tennodai, Tsukuba, Ibaraki 305-8577, Japan}

\author{Zarija Luki\'c}
\affil{Computational Cosmology Center, Lawrence Berkeley National Laboratory, CA 94720, USA}

\begin{abstract} 
Ly$\alpha$ intensity mapping is emerging as a new probe of faint galaxies consisting the cosmic web that elude traditional surveys. However, the resonant nature of Ly$\alpha$ radiative transfer complicates the interpretation of observed data. In this study, we develop a fast and accurate analytic prescription for computing the Ly$\alpha$ intensity field on Mpc scales in the post-reionization Universe. Motivated by insights from Monte Carlo radiative transfer (MCRT) experiments, we exploit the fact that in a highly ionized intergalactic medium (IGM) with negligible damping-wing opacity, cosmological redshifting quickly drives Ly$\alpha$ photons out of resonance, terminating the scattering process and simplifying their large-scale behavior. Photons emitted blueward of the Ly$\alpha$ line center tend to scatter on a thin, nearly spherical surface of last scattering, with a radius determined by the redshifting distance to resonance. Based on this behavior, we derive closed-form expressions for the scattered emissivity and projected surface brightness that depend only on the source spectrum, the HI density, and the peculiar velocity field. When applied to a source in a realistically simulated IGM at $z = 3$, our model shows mild discrepancies with MCRT results within a physical Mpc of the host halo, where strong gravitational infall redistributes the scattered photons, but achieves better than 5\% accuracy beyond that distance in angle-averaged radial surface brightness profile. Our prescription offers a computationally efficient alternative to MCRT for forward-modeling Ly$\alpha$ intensity maps from cosmological simulations, enabling the inference of underlying cosmological and astrophysical parameters from future observations.

\vspace{1em}
\end{abstract}

% \vspace{0.5em}

%cosmology: theory < Cosmology, radiative transfer < Physical Data and Processes, (galaxies:) intergalactic medium < Galaxies, software: simulations < Software

% \begin{keyword}
%  Cosmology
% \end{keyword}

\section{Introduction}

Mapping the large-scale distribution of matter in the Universe provides critical insights into the formation and evolution of cosmic structures. In the present-day Universe, most matter resides in the cosmic web—a network of filamentary structures formed through the gravitational evolution of primordial overdensities \citep{1996Natur.380..603B}. Among the expanding set of observational techniques, Ly$\alpha$ intensity mapping (LIM) is emerging as a promising method for probing the faint galaxy populations that trace these filaments but lie beyond the reach of traditional galaxy surveys \citep[e.g.,][]{2016MNRAS.462.1961S,2021A&A...650A..98W,2022ApJS..262...38L,2024MNRAS.535..826R}. Theoretical studies have shown that LIM can capture additional astrophysical and cosmological information by integrating Ly$\alpha$ emission from both resolved and unresolved sources across large cosmic volumes \citep[e.g.,][]{2011ApJ...739...62Z,2017ApJ...848...52H,2020MNRAS.494.5439E,2023MNRAS.523.5248B,2023NatAs...7.1390M,2024arXiv241000450T,2025ApJ...984...55L}.

Ly$\alpha$ photons originate from the recombination of ionized hydrogen and the collisional excitation of hydrogen atoms in dense environments inside energetic sources such as star-forming galaxies and active galactic nuclei \citep[see e.g.,][ for review]{2014PASA...31...40D}. These photons can undergo resonant scattering by neutral hydrogen in the intergalactic medium (IGM), producing extended emission that traces the underlying cosmic web. Recent observations suggest that the sky is permeated by diffuse Ly$\alpha$ emission from these mechanisms, with scattering occurring on both galactic and cosmological scales \citep{2018Natur.562..229W}.

Extended Ly$\alpha$ emission is observed on a wide range of spatial scales. Emission on scales of 1–10 kpc is referred to as Ly$\alpha$ halos (LAHs) and has been detected through both stacking analyses \citep{2004AJ....128.2073H,2011ApJ...736..160S,2012MNRAS.425..878M,2014MNRAS.442..110M,2016MNRAS.457.2318M,2017ApJ...837..172X,2022ApJ...929...90L} and deep spectroscopic observations of individual targets \citep{2016A&A...587A..98W,2017A&A...608A...8L}. Ly$\alpha$ blobs, which are much larger than typical LAHs, span $\sim$10 to 100 physical kpc and typically exhibit Ly$\alpha$ surface brightness of $\sim 10^{-18}~{\rm erg}~{\rm s}^{-1}~{\rm cm}^{-2}~{\rm arcsec}^{-2}$ \citep{1987ApJ...319L..39M,2000ApJ...532..170S,2004AJ....128..569M,2011MNRAS.410L..13M,2018PASJ...70S..15S}.

Ly$\alpha$ emission tends to be more extended around more energetic sources and in overdense environments. Enormous Ly$\alpha$ nebulae, spanning hundreds of physical kiloparsecs, are frequently observed around quasars \citep{2014Natur.506...63C,2015Sci...348..779H,2016ApJ...831...39B}. Cross-correlation Ly$\alpha$ intensity mapping has revealed that Ly$\alpha$ emission around these quasars extends to several megaparsecs \citep{2016MNRAS.457.3541C,2018MNRAS.481.1320C,2022ApJS..262...38L}. A direct detection of excess Ly$\alpha$ intensity around a bright quasar has also been reported at $z=3.1$ \citep{2019Sci...366...97U}. Ly$\alpha$ emitters (LAEs) also are found to exhibit spatial correlations with Ly$\alpha$ intensity on megaparsec scales at $z \sim 2$–6 \citep{2021ApJ...916...22K,2022ApJ...931...97K}. 

Mapping Ly$\alpha$ filaments has been technically challenging due to their large spatial extents and extremely low surface brightness \citep[$\lesssim 10^{-20}~{\rm erg}~{\rm s}^{-1}~{\rm cm}^{-2}~{\rm arcsec}^{-2}$;][]{2018MNRAS.475.3854G}. However, recent technological advances are beginning to enable Ly$\alpha$ intensity mapping with sufficiently wide fields of view and high sensitivity. The Multi Unit Spectroscopic Explorer \citep[MUSE;][]{2010SPIE.7735E..08B} has successfully detected diffuse Ly$\alpha$ emission tracing cosmic web structures in recent years \citep[e.g.,][]{2021A&A...647A.107B,2025ApJ...980L..43T}. Notably, the MIRACLES survey (Mapping of Ionizing RAdiation on the Cosmic web with Ly$\alpha$ Emission and Shadow; PI: Y. Matsuda) is using the Hyper Suprime-Cam (HSC) instrument on the Subaru telescope to measure the diffuse Ly$\alpha$ emission at $z = 3.09$ and is already producing extended Ly$\alpha$ maps across several megaparsecs (Mawatari et al., in preparation).

Given this observational progress, it is becoming increasingly crucial to improve our theoretical understanding of Ly$\alpha$ radiative transfer in the IGM to accurately interpret upcoming data. Typically, resonant scattering of Ly$\alpha$ photons is modeled by post-processing snapshots from cosmological hydrodynamic simulations using Monte Carlo radiative transfer (MCRT) methods \citep[e.g.,][]{2011ApJ...739...62Z,2012MNRAS.424..884Y,2012ApJ...754..118Y,
2014MNRAS.440..776Y,2015MNRAS.449.4336S,2018A&A...614A..31B,2020A&A...635A.154M,2021MNRAS.506.5129B,2024MNRAS.535.1979M,2025JCAP...08..080A}. Recent simulations indicate that resonantly scattered Ly$\alpha$ radiation is likely a major contributor to the large-scale correlation between galaxies and Ly$\alpha$ surface brightness across the sky \citep[e.g.,][]{2023MNRAS.523.5248B}.
%1968ApJ...153..783A,2002ApJ...567..922A,2002ApJ...578...33Z,2006MNRAS.367..979H,2010MNRAS.403..870B,2011A&A...531A..12S,2020ApJS..250....9S,

%2006ApJ...645..792T,2007ApJ...657L..69L,2010ApJ...708.1048K,

%2012ApJ...754..118Y, 2014MNRAS.440..776Y

In our recent work \citep{2022ApJ...931..126P}, we investigated Ly$\alpha$ scattering in the context of a reionizing IGM at $z \sim 7$ and found that the process can be substantially simplified in ionized regions due to negligible damping-wing opacity. We showed that photons emitted at frequencies blueward of Ly$\alpha$ are scattered only after cosmologically redshifting into near-resonance, forming thin, nearly spherical scattering surfaces on IGM scales. The resulting spectral energy distribution (SED) of the scattered light was consistent with that of a spherical source projected onto the sky. In this work, we translate this physical picture into analytic expressions applicable to the fully ionized, post-reionization Universe and validate them by comparing with MCRT simulations in a realistically simulated IGM. Our goal is to develop a computationally efficient, semi-analytic prescription that would significantly reduce the computational cost of generating Ly$\alpha$ intensity maps across many mock IGM datasets, enabling systematic exploration of their parametric dependence in preparation for upcoming surveys. Our analytic model for the post-reionization IGM would complement other existing analytic solutions derived for individual galaxies \citep{2025MNRAS.537.1646N} and the neutral IGM \citep{1999ApJ...524..527L,2025MNRAS.541..179S}.

The rest of this paper is organized as follows. In Section~\ref{sec:insight}, we revisit our previous insight into Ly$\alpha$ scattering in the ionized IGM using our MCRT simulation code in an idealized setup and derive analytic expressions for the scattered Ly$\alpha$ intensity map from a given source SED. In Section~\ref{sec:test}, we validate our analytic model by comparing its predictions against MCRT simulations in a cosmological hydrodynamic simulation. In Section~\ref{sec:summary}, we summarize and discuss our results. We assume a cosmology with $h = 0.678$, $\Omega_{\rm M} = 0.307$, and $\Omega_{\rm b} = 0.048$, consistent with measurements from the Planck satellite \citep{2020A&A...641A...6P}. Comoving and physics megaparsecs are denoted as Mpc and pMpc, respectively.

\section{Insights from Ly$\boldsymbol{\alpha}$ Radiative Transfer Simulations}\label{sec:insight}

\begin{figure*}
  \begin{center}
    \includegraphics[scale=0.8]{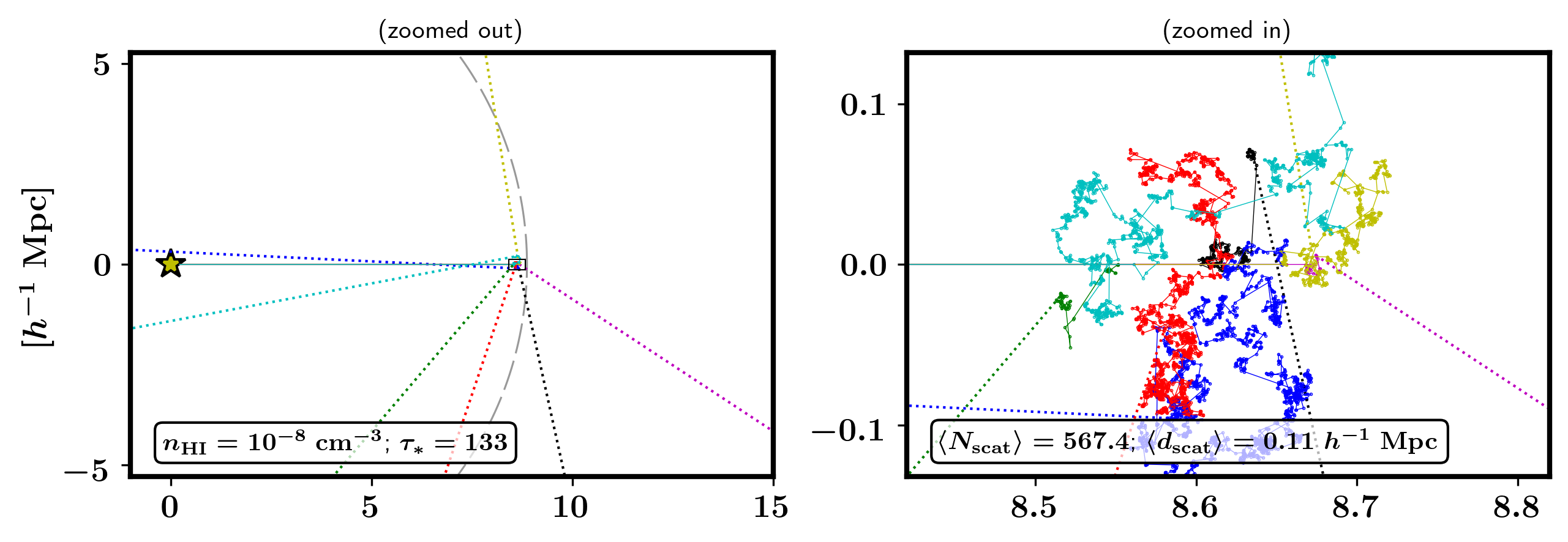}
    \includegraphics[scale=0.8]{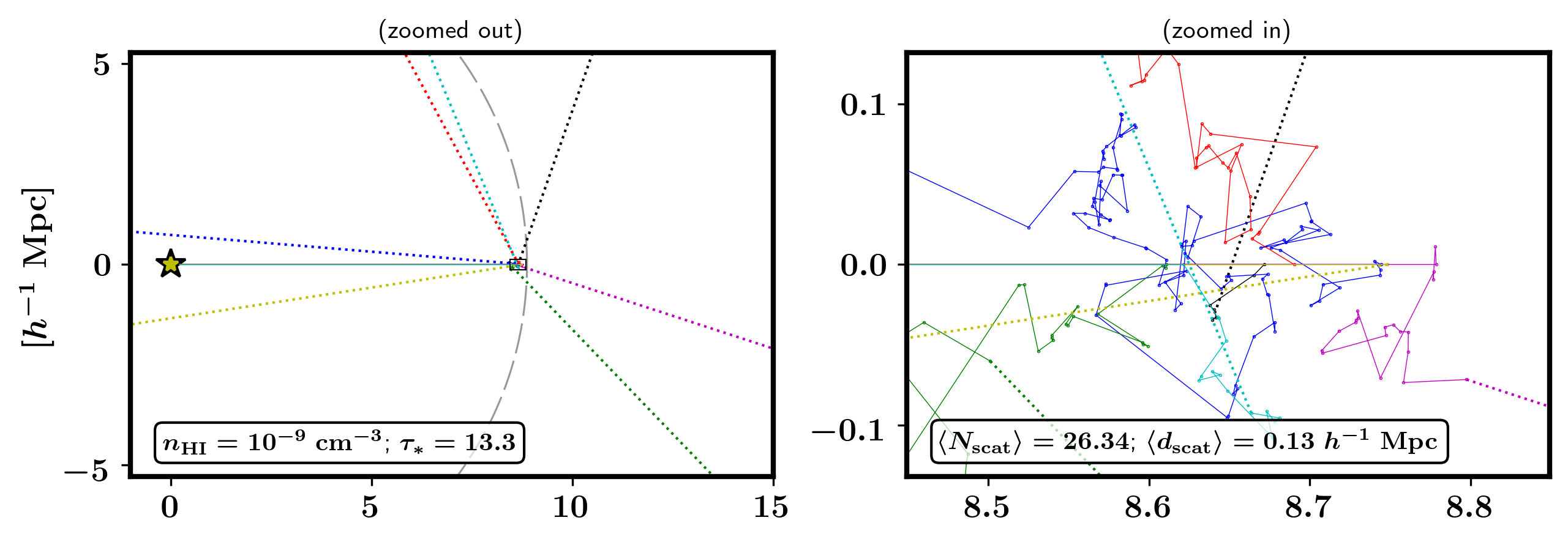}
    \includegraphics[scale=0.8]{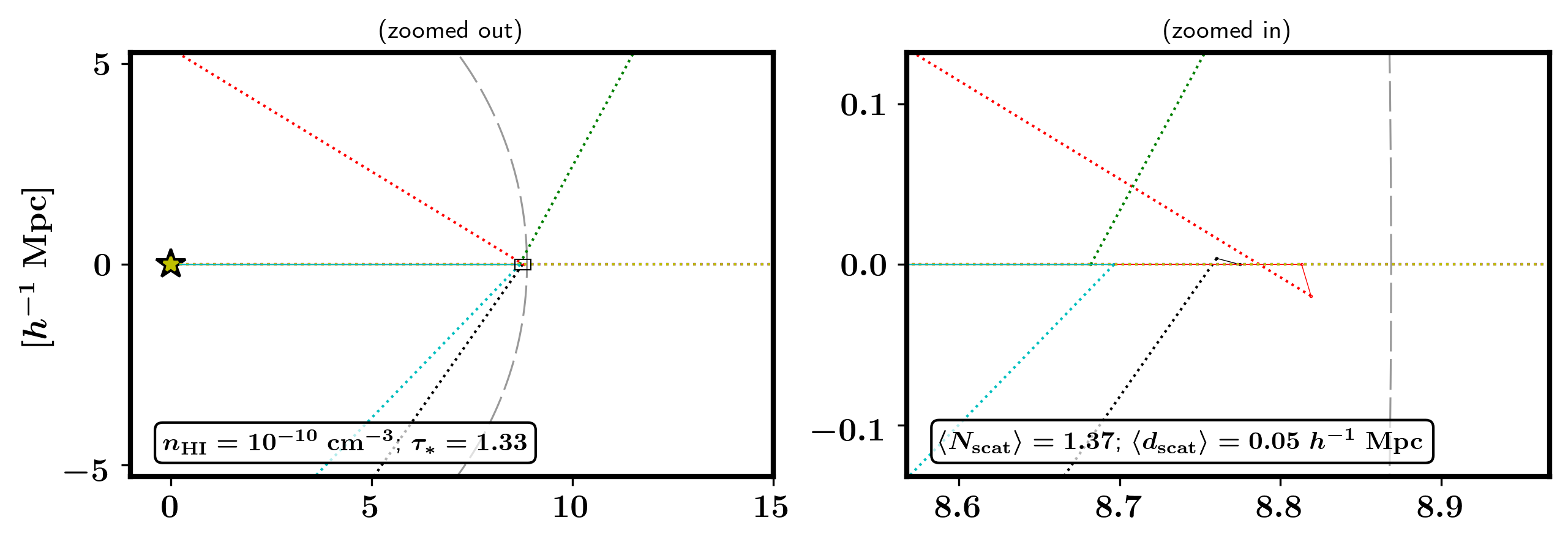}
  \caption{ Trajectories of photons initially emitted 4.06Å blueward of Ly$\alpha$ ($v_{\alpha,{\rm i}} = -1000~{\rm km}{\rm s}^{-1}$; $\Delta\lambda_{\alpha,{\rm i}} = -4.06~{\rm Å}$) in the $+x$ direction, projected onto the $xy$ plane. The IGM is assumed to be static and to have uniform neutral hydrogen densities of $10^{-8}$, $10^{-9}$, and $10^{-10}~{\rm cm}^{-3}$, corresponding to Ly$\alpha$ optical depths (Eq.~\ref{eq:tau_star}) of $\tau_* = 133$, 13.3, and 1.33 in the upper, middle, and lower panels, respectively. The star symbol marks the photon source, and the black dashed circle in the left panels indicates the last-scattering distance $r_{\rm s}$ as described in Section~\ref{sec:LSS}. For each HI density, the trajectories of seven photons are shown as solid and dotted lines, with the dotted lines marking the final paths taken by the photons as they escape. The right panels provide zoomed-in views of the regions where most scatterings occur. In the lower-left corner of each right panel, we report the average number of scatterings, $\left<N_{\rm scat}\right>$, and the average separation between the first and last scattering, $\left<d_{\rm scat}\right>$.}
   \label{fig:trajectories}
  \end{center}
\end{figure*} 

In our previous work \citep{2022ApJ...931..126P}, we developed and validated a MCRT simulation code to model the propagation of Ly$\alpha$ photons in the intergalactic medium (IGM) and presented a detailed analysis of their scattering processes on cosmological scales. In this study, we revisit those findings by running the MCRT simulations in simplified setups to derive and justify our analytic model for the Ly$\alpha$ intensity map of scattered photons. Throughout this work, we neglect absorption by dust, as the high-redshift IGM is expected to be largely dust-poor. For validation, we track all scattering events occurring during the MCRT runs.

\subsection{Deterministic Scattering Location of Ly$\alpha$ Photons} \label{sec:scattering}

Figure~\ref{fig:trajectories} shows the trajectories of photons initially emitted toward a same direction at a wavelength slightly bluer than the Ly$\alpha$ resonance ($\lambda_\alpha=1215.67$ Å) by $\Delta \lambda_{\alpha, {\rm i}}\equiv \lambda_{\rm i} -\lambda_\alpha = -4.06~{\rm Å}$ in wavelength or $v_{\alpha,{\rm i}} \equiv c[\Delta \lambda_{\alpha,{\rm i}}/\lambda_\alpha] = -1000~{\rm km}~{\rm s}^{-1}$ in velocity. In this experiment, we assume a static monochromatic source at $z=3$ emitting photons into a static IGM with a uniform HI density of $n_{\rm HI} = 10^{-8}$, $10^{-9}$, and $10^{-10}~{\rm cm}^{-3}$ at a temperature of $T_{\rm IGM} =$ 10,000 K, mimicking typical IGM conditions in the post-reionization Universe.  The cosmic mean hydrogen density (including both ionized and neutral components) at this redshift is $\bar{n}_{\rm H}=1.22\times10^{-5}~{\rm cm}^{-3}$. A value of $n_{\rm HI} = 10^{-10}~{\rm cm}^{-3}$ corresponds to a typical IGM near the cosmic mean density, with a neutral fraction of approximately $10^{-5}$, while $n_{\rm HI} = 10^{-9}$ and $10^{-8}~{\rm cm}^{-3}$ represent increasingly overdense regions.

The damping-wing opacity of Ly$\alpha$ becomes significant only when the neutral fraction of the IGM approaches unity \citep[e.g.,][]{1998ApJ...501...15M}, whereas the environments we consider here are post-reionization, with neutral fractions several orders of magnitude lower. As a result, most photons propagate freely until they reach the {\it Sobolev resonance point} \citep{1960mes..book.....S}, where they redshift into resonance and begin scattering. This first scattering occurs at a distance of approximately $8.8~h^{-1}~{\rm Mpc}$ from the source, marked by the gray dashed circle in Figure~\ref{fig:trajectories}. After the initial scattering, each photon undergoes multiple additional scatterings within a compact region of $\sim 0.1~h^{-1}~{\rm Mpc}$ before escaping to infinity, as illustrated in the right-hand zoom-in panels. On larger scales ($\gtrsim 1~{\rm Mpc}$; left panels), the photon's trajectory effectively resembles a single scattering event, after which it escapes in a random direction—resulting in an isotropic light source located at the resonance point, largely independent of local density.

Scattering begins when a photon, initially emitted at a wavelength bluer than the Ly$\alpha$ resonance, is cosmologically redshifted to near-resonance. At $z\gtrsim 2$, where dark energy is negligible, the cosmic expansion rate can be written as
\bea \label{eq:H}
H(z) &\approx& \left[\frac{301~{\rm km}~{\rm s}^{-1}}{\rm pMpc}\right] \left[\frac{h}{0.678}\right]\left[\frac{\Omega_{\rm M}}{0.307}\right]^{0.5} \left[\frac{1+z}{4}\right]^{1.5}
% \nonumber\\
% &=& \left[\frac{113~{\rm km}~{\rm s}^{-1}}{h^{-1}~{\rm Mpc}}\right] \left[\frac{\Omega_{\rm M}}{0.307}\right]^{0.5} \left[\frac{1+z}{4}\right]^{0.5}
.
\eea
In our experiment for the $z=3$ IGM, photons emitted at $v_{\alpha,{\rm i}} = -1000~{\rm km}~{\rm s}^{-1}$ redshift to the resonance after traveling a comoving distance of $-v_{\alpha,{\rm i}}/(aH) \approx  8.8~h^{-1}~{\rm Mpc}$. Here, we refer to this as the scattering distance.

Thermal motion of atoms can broaden the cross section in wavelength and cause the scattering to occurs even when the wavelength is not exactly at the resonance (1215.67 Å). However, the thermal broadening occurs at scales below Mpc. The thermal velocity of $10^4~{\rm K}$ gas is
\bea \label{eq:v_th}
v_{\rm th} \equiv \sqrt{\frac{2 k_{\rm B} T_{\rm IGM}}{m_{\rm p}}} \approx 12.9~[{\rm km}~{\rm s}^{-1}] \left[\frac{T}{10^4~{\rm K}}\right]^{0.5},
\eea
where $k_{\rm B} = 1.38\times 10^{-16}~{\rm erg}\cdot{\rm K}$ is the Boltzmann constant and $m_{\rm p} = 1.67\times 10^{-24}~{\rm g}$ is the proton mass. This thermal velocity corresponds to a {\it Sobolev length} of $v_{\rm th}/(aH) \approx 0.11~h^{-1}~{\rm Mpc}$ or wavelength of $\lambda_\alpha v_{\rm th}/c=0.052~{\rm Å}$ at $z = 3$, within which scattering can occur. Thus, the scattering process is confined to scales much smaller than the total photon travel distance ($\sim 10~{\rm Mpc}$). Increasing the HI density increases the number of scatterings, but the large-scale behavior of the photon trajectories remains essentially unchanged.

%This explains why the photon trajectory appears as a single directional change at Mpc scales in the left panels of Figure~\ref{fig:trajectories}. The right panels of Figure~\ref{fig:trajectories} reveal that photons undergo nearly all their scatterings within a compact region of $\sim 0.1~{\rm Mpc}$. 

When $n_{\rm HI}\lesssim 10^{-10}~{\rm cm}^{-3}$, a significant fraction of photons escapes the system unscattered due to the low opacity. The yellow trajectory in the lower panels of Figure~\ref{fig:trajectories} illustrates one of such cases. These photons would appear to observers as a part of the unscattered radiation directly coming from the source. The fraction of unscattered photons can be obtained as $e^{-\tau_*}$, where $\tau_*$ is calculated from integrating the Ly$\alpha$ opacity along a straight line to infinity:
\bea \label{eq:tau_star}
\tau_* &\approx& \left[\frac{n_{\rm HI}}{0.75\times10^{-10}~{\rm cm}^{-3}}\right] \nonumber\\
&&\times \left[\frac{h}{0.678}\right]^{-1} \left[\frac{\Omega_{\rm M}}{0.307}\right]^{-0.5} \left[\frac{1+z}{4} \right]^{-1.5},
\eea
which assumes the Hubble expansion rate is in the matter dominated regime ($[1+z]^3\gg 1$), and the fraction of scattered photons is given by $1 - e^{-\tau_*}$. The detailed derivation for the above equation is given in the Appendix. In a non-uniform IGM (i.e., $n_{\rm HI} = n_{\rm HI}(\mathbf{r})$), Equation~(\ref{eq:tau_star}) remains valid due to the short Sobolev length—the dominant contribution to $\tau_*(\mathbf{r})$ comes from the local value of $n_{\rm HI}(\mathbf{r})$ on Mpc scales. Over a broad range of $n_{\rm HI}$, from $10^{-9}$ to $10^{-5}~{\rm cm}^{-3}$, where the resonance opacity is high enough to scatter nearly all the blue-ward photons ($\tau_*\gg1$) yet the damping-wing opacity remains negligible, the scattered light becomes largely insensitive to the exact value of $n_{\rm HI}$, as shown in Figure~\ref{fig:trajectories}.

\begin{figure*}
  \begin{center}
    \includegraphics[scale=0.605]{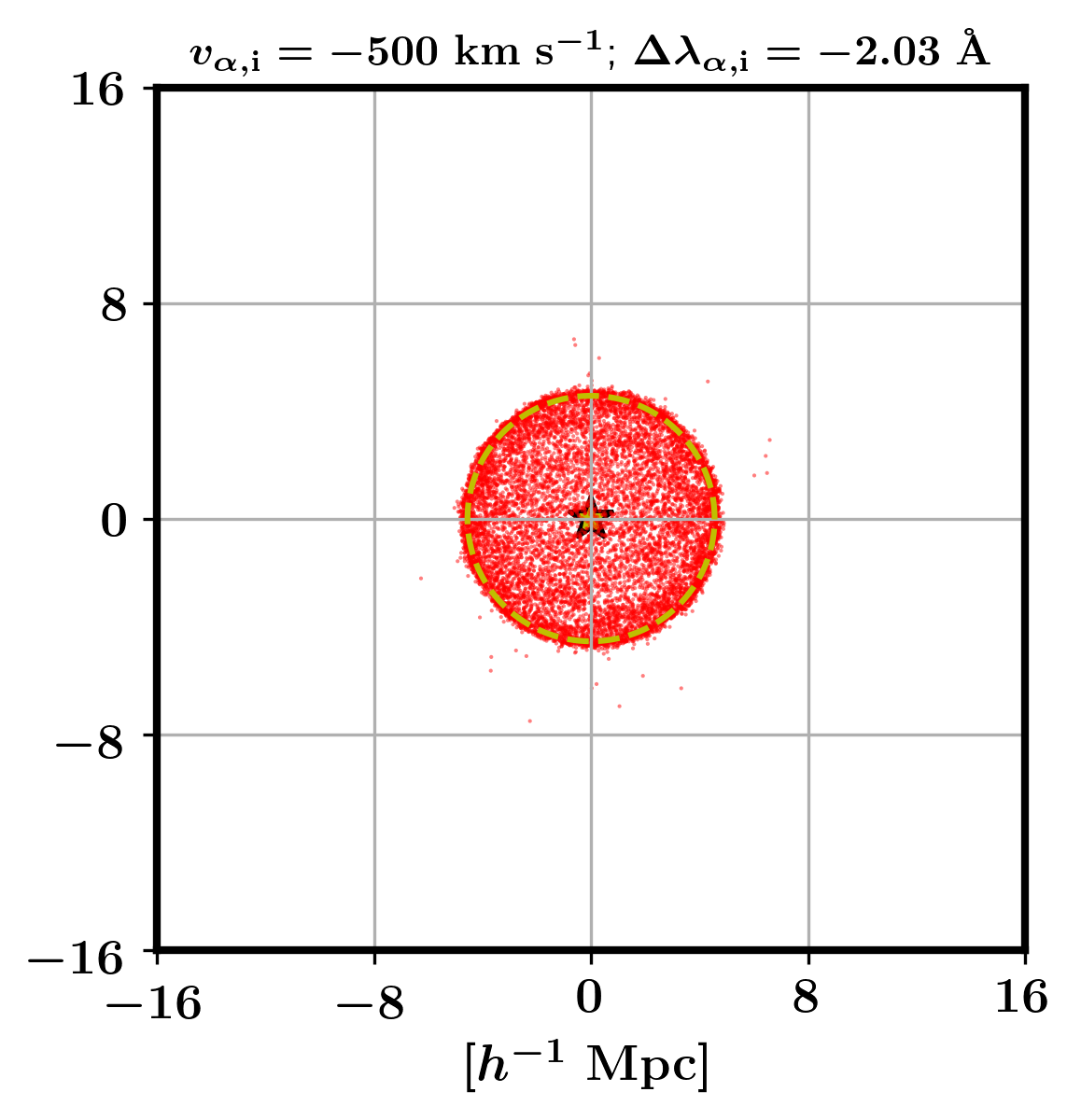}
    \includegraphics[scale=0.605]{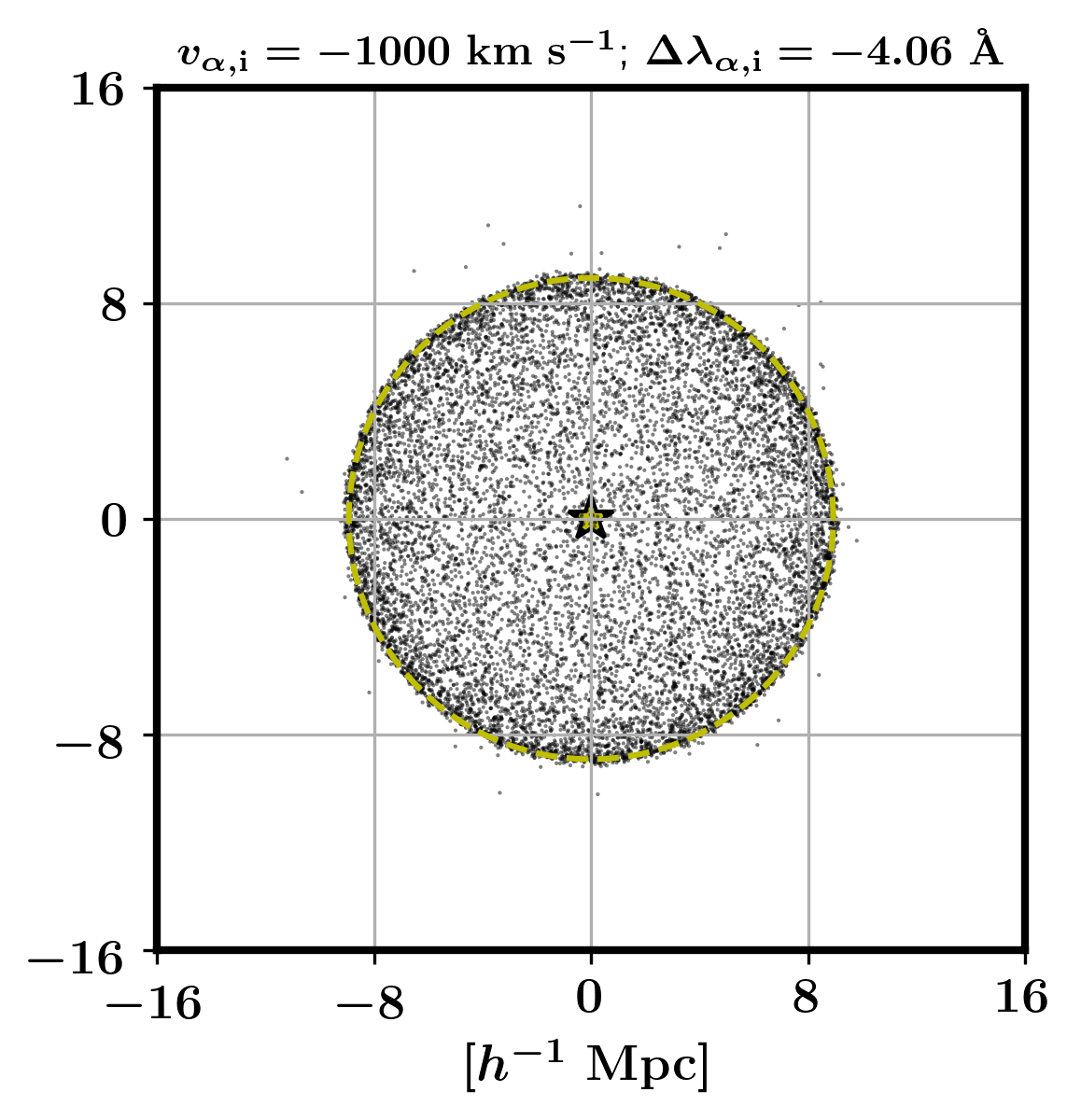}
    \includegraphics[scale=0.605]{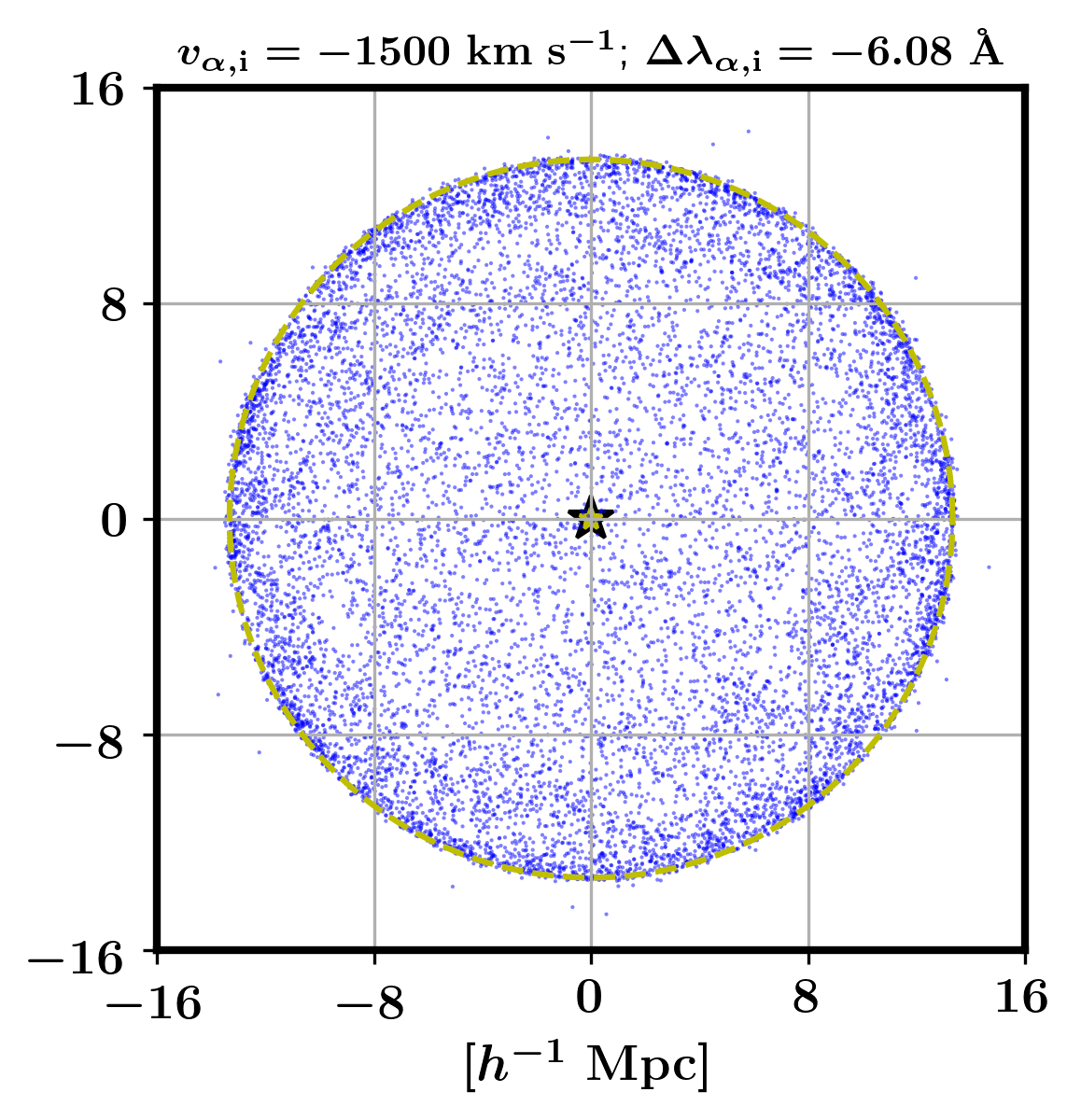}
  \caption{Location of last scattering events projected on the $xy$ plane for photons emitted isotropically in the static, uniform IGM with $n_{\rm HI}=10^{-9}~{\rm cm}^{-3}$. The results are shown for initial velocity offset of $v_{\alpha,{\rm i}}=-500$ (left), $-1000$ (middle), and $-1500~{\rm km}~{\rm s}^{-1}$ (right panel), corresponding to wavelength offset of $\Delta \lambda_{\alpha,i}=-2.03,~ -4.06$, and $-6.08~{\rm Å}$, respectively. The dashed circle in each panel indicates the scattering distance $r_{\rm s}$ from the source corresponding to the respective initial wavelength at emission given by Equation~(\ref{eq:r_s}). }
   \label{fig:LSS}
  \end{center}
\end{figure*} 

\subsection{Scattering Surface for Each Emission Wavelength} \label{sec:LSS}

The experiment presented above can be generalized to a source that emits photons isotropically at a given frequency. In this case, the scattering locations of the photons would be distributed over a spherical surface whose radius is the scattering distance. Figure~\ref{fig:LSS} shows this spherical distribution of last-scattering locations (projected on the $xy$ plane) for the photons that are emitted into uniformly random directions at three different velocity offsets $v_{\alpha,{\rm i}}=-500,~-1000,$ and $-1500~{\rm km}~{\rm s}^{-1}$ in the uniform IGM with $n_{\rm HI}=10^{-9}~{\rm cm}^{-3}$, corresponding to $\tau_*\approx13.3$.

In more general cases where the IGM has a peculiar velocity, $\mathbf{v}_{\rm pe}$, the scattering distance is given by
\bea \label{eq:r_s}
r_{\rm s} \equiv -\frac{v_{\alpha,{\rm i}} + v_{{\rm pe},r}(\mathbf{r})}{aH(z)} \, ,
\eea
where $a=[1+z]^{-1}$ is the scale factor and $v_{{\rm pe},r}\equiv \hat{\mathbf{r}}\cdot \mathbf{v}_{\rm pe}$ is the radial component of the peculiar velocity, accounting for the Doppler shift of photon wavelengths in the rest frame of the IGM. Here, $\mathbf{r}$ denotes the location relative to the source in the comoving coordinate. The peculiar velocity introduces slight deviations from a perfect sphere in the scattering surface. However, the surface remains nearly spherical since the Hubble flow dominates on cosmological scales. A more realistic scattering surface from a cosmological simulation can be found in \cite{2022ApJ...931..126P} (see, e.g., their Fig. 7).

As long as the radial gradient of the radial peculiar velocity, $\text{d}v_{{\rm pe},r}/\text{d}r$, remains smaller than the Hubble expansion rate, $H(z)$—which is typically the case on IGM scales—the scattering surface expands monotonically in all directions as $v_{\alpha,{\rm i}}$ (or equivalently, $\lambda_{\rm i}$) decreases. This implies that every point in space, $\mathbf{r}$, corresponds to a unique initial wavelength $\lambda_{\rm i}$ whose scattering surface encompasses that location. Consequently, the scattered radiation emissivity, $\epsilon_{\rm scat}(\mathbf{r})$, at each location $\mathbf{r} = r_{\rm s} \hat{r}$ is solely determined by the source luminosity at the specific $v_{\alpha,{\rm i}}$ corresponding to $r_{\rm s} = |\mathbf{r}|$ via Equation~(\ref{eq:r_s}). This allows us to derive an analytic expression for $\epsilon_{\rm scat}$ below.

\subsection{Analytic Model for Ly$\alpha$ Intensity Mapping}

The existence of a unique wavelength that contributes scattered radiation to a given location allows us to construct a closed-form analytic expression for the Ly$\alpha$ intensity map, providing a computationally efficient way to generate Ly$\alpha$ intensity maps. $\epsilon_{\rm scat}$ can be obtained by spreading the luminosity at the the corresponding wavelength over the scattering surface. For a source with luminosity per wavelength $L_\lambda(\lambda_{\rm i})\equiv \text{d}L/\text{d}\lambda$ at the wavelength $\lambda_{\rm i}$, we have:
\bea \label{eq:e_scat}
4\pi r^2 a^3\epsilon_{\rm scat}(\mathbf{r}) \text{d}r  = [1-e^{-\tau_*}] L_\lambda(\lambda_{\rm i}) \text{d}\lambda_{\rm i} \, ,
\eea
where $\lambda_{\rm i}=\lambda_\alpha [1+v_{\alpha,{\rm i}}/c]$ is the wavelength that will be at resonance at the position $\mathbf{r}$ and therefore satisfies Equation~(\ref{eq:r_s}) for $r_{\rm s}=|\mathbf{r}|$.

\begin{figure*}
  \begin{center}
    \includegraphics[scale=0.55]{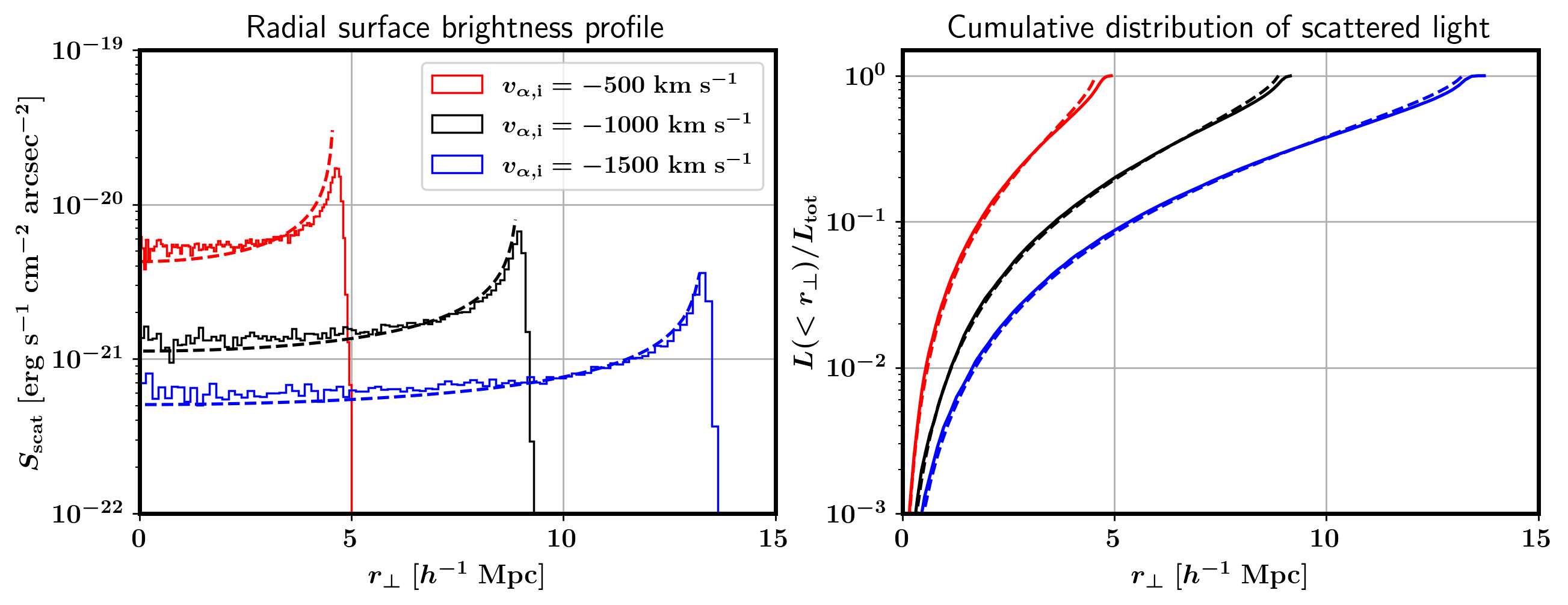}
  \caption{Radial surface brightness profile (left) and cumulative distribution (right) of the scattered Ly$\alpha$ photons as a function of projected distance from an isotropic monochromatic source. We assume a source luminosity of $L_{\rm tot}=10^{44}~{\rm erg}~{\rm s}^{-1}$ and an HI density of $n_{\rm HI}=10^{-9}~{\rm cm}^{-3}$. Simulation results are shown as histograms outlined by solid lines, while the analytic model from Equation~(\ref{eq:SBmono}) is shown as dashed lines. The upper, middle, and lower curves correspond to $v_{\alpha,{\rm i}}=-500$, $-1000$, and $-1500~{\rm km}~{\rm s}^{-1}$, respectively. }
   \label{fig:rPjt}
  \end{center}
\end{figure*} 

From this, we derive the scattering emissivity as
\bea \label{eq:e}
    \epsilon_{\rm scat}(\mathbf{r}) &=& \frac{L_\lambda(\lambda_{\rm i})}{4\pi r^2 a^2} \frac{\text{d}\lambda_{\rm i}}{a\text{d}r} [1-e^{-\tau_*}] \nonumber \\
    &=& \frac{L_\lambda(\lambda_{\rm i})}{4\pi r^2 a^2} \left[ \frac{\lambda_\alpha}{c} \right]
    \left[ H(z) + \frac{\text{d}v_{{\rm pe},r}}{a\text{d}r} (\mathbf{r}) \right] [1-e^{-\tau_*}],~~~~
\eea
where we assume the Hubble rate $H(z)$ remains approximately constant on the scales of interest. 

% precompute and store the functional form of $\epsilon_{\rm scat}$ (excluding the $1 - e^{-\tau_*}$ factor) and reuse it for sources with similar properties—such as halo mass or star formation rate.

The observed surface brightness, $S_{\rm Ly\alpha}$, is obtained by projecting $\epsilon_{\rm scat}(\bold{r})$ onto the sky plane and calculating the resulting intensity reaching the observer. Let $\text{d}A$ be the projected area around the source and $\text{d}\Omega$ the solid angle it subtends as seen by the observer. Then we can write 
\bea
S_{\rm scat}(\mathbf{r}_\perp) \text{d}\Omega  &=&  \frac{\text{d}A}{4\pi D_{\rm L}^2}  \int^{r_{\parallel,{\rm u}}}_{r_{\parallel,{\rm l}}} 
\epsilon_{\rm scat}(\mathbf{r})\, a\, \text{d}r_\parallel \, ,
\eea
where $D_{\rm L}$ is the luminosity distance, and $r_\parallel$ and $\mathbf{r}_\perp$ are projections of $\mathbf{r}$ onto the sight line and the sky plane. Here, we assume that the emissivity of the scattered light is isotropic, as scattered photons are equally likely to escape to any direction once they drift out of the resonance core. This is justified because the post-reionization IGM has negligible damping-wing opacity. We confirm this assumption with our simulation results.

The line-of-sight integration range $[r_{\parallel,{\rm l}},r_{\parallel,{\rm u}}]$ depends of the specifics of the surface brightness survey, such as the wavelength coverage of the filter in the case of narrow-band imaging. 
We can further derive:
\bea \label{eq:S}
S_{\rm scat}(\mathbf{r}_\perp) &=& \frac{D_{\rm A}^2}{4\pi D_{\rm L}^2}  \int^{r_{\parallel,{\rm u}}}_{r_{\parallel,{\rm l}}} \epsilon_{\rm scat}(\mathbf{r})\,a\,\text{d}r_\parallel~[{\rm sr}^{-1}]\nonumber \\
&=& \frac{a^4}{4\pi}  \int^{r_{\parallel,{\rm u}}}_{r_{\parallel,{\rm l}}} \epsilon_{\rm scat}(\mathbf{r}) \,a\,\text{d}r_\parallel~[{\rm sr}^{-1}] \, ,
\eea
where $D_{\rm A}$ is the angular diameter distance, and the factor $a^4$ accounts for the cosmological surface brightness dimming.

In the monochromatic source case from Section~\ref{sec:LSS}, the luminosity can be written as $L_\lambda(\lambda)=L_{\rm tot} \delta^{\rm D}(\lambda-\lambda_{\rm i})$, where $\delta^{\rm D}$ is the Dirac Delta function. The peculiar velocity $\text{d}v_{{\rm pe},r}/\text{d}r$ is set to zero as we assumed the IGM is static. If we further assume the IGM is highly opaque at the resonance (i.e., $\tau_*\gg1$) and the line-of-sight integration range is wide enough to enclose all scattered radiation (i.e., $r_{\parallel,{\rm u}}=\infty$ and $r_{\parallel,{\rm l}}=-\infty$), we have
\bea
S_{\rm scat} (r_\perp) &=& 
\frac{a^5}{4\pi}  \int \frac{L_{\rm tot}}{4\pi r_{\rm s}^2 a^3} \frac{\text{d}\lambda}{\text{d}r} \delta^{\rm D}(\lambda-\lambda_{\rm i}) \text{d}r_\parallel \quad [{\rm sr}^{-1}]
\nonumber \\
&=& \frac{a^2 L_{\rm tot}}{16\pi^2 r_{\rm s}^2}
\int \delta^{\rm D}(r-r_{\rm s}) \text{d}r_\parallel \quad [{\rm sr}^{-1}].
\eea
Here, we use the fact that $\text{d}r_\parallel = [r/r_\parallel]\text{d}r = r\,\text{d}r / \sqrt{r^2 - r_\perp^2}$ when $r_\perp$ is fixed, and that $r = r_{\rm s}$ is satisfied twice at $r_\parallel = \pm \sqrt{r_{\rm s}^2 - r_\perp^2}$ within the integration. This gives:
\bea \label{eq:SBmono}
S_{\rm scat} (r_\perp) &=& \frac{a^2 L_{\rm tot}}{16\pi^2 r_{\rm s}^2} \frac{2r_{\rm s}}{\sqrt{r_{\rm s}^2-r^2_\perp}} \quad [{\rm sr}^{-1}] \nonumber \\
&=& \frac{a^2 L_{\rm tot}}{8\pi^2 r_{\rm s}^2\sqrt{1-[r_\perp/r_{\rm s}]^2}} \quad [{\rm sr}^{-1}].
\eea 
This analytic expression agrees well with the simulation results shown in Figure~\ref{fig:rPjt}, both in terms of the intensity profile and its cumulative distribution as a function of projected distance from the source.\footnote{Note that an additional conversion factor is needed to express intensity in ${\rm arcsec}^{-2}$.} In this comparison, we assumed $L_{\rm tot}=10^{44}~{\rm erg}~{\rm s}^{-1}$ and $n_{\rm HI}=10^{-9}~{\rm cm}^{-3}$. Agreement for the monochromatic source case naturally extends to general multi-wavelength sources, as they can be expressed as a superposition of monochromatic components.

\subsubsection{Computational cost analysis}

The majority of the computational cost in the proposed analytical model arises from evaluating Equation~(\ref{eq:e}) at each location $\mathbf{r}$ in a 3D volume. Each evaluation requires a constant number of arithmetic operations for a given spectral shape of the Ly$\alpha$ source, $L_\lambda(\lambda_{\rm i})$, including the computation of ${\rm d}v_{{\rm pe},r}/{\rm d}r$ and $\tau_*$. In practical applications, this process must be repeated for all sources contributing to the intensity field. Consequently, the total computational cost of calculating the intensity field scales linearly with the product of the number of sources and the spatial grid size on which the intensity is evaluated, $\mathcal{O}(\mathcal{N}_{\rm source}\mathcal{N}_{\rm cell})$.

While the computational cost is generally expected to be much lower than that of full numerical simulations, it can increase steeply when accounting for contributions from the numerous faint sources that provide a significant cumulative signal, or when modeling the intensity map over large scales ($\gtrsim 100~{\rm Mpc}$) with high spatial resolution ($\lesssim 0.1~{\rm Mpc}$). Possible strategies to mitigate this scaling include merging sources that are separated by less than the cell size or resolution limit of the target survey, and skipping emissivity calculations beyond the distance at which contributions become negligible. 

We note that $\tau_*$ can be reused for other sources by applying appropriate coordinate shifts corresponding to different source locations. However, $\text{d}v_{{\rm pe},r}/\text{d}r$ must be recalculated for each source, as it depends on both the source position and its peculiar velocity. One possible approximation is to adopt an average peculiar velocity gradient for sources with similar halo masses (or other available properties that correlate with halo mass) to compute the functional form of $\epsilon_{\rm scat}$ (excluding the $1 - e^{-\tau_*}$ factor) and reuse it for applicable sources. This approach is motivated by the finding of \cite{2021ApJ...922..263P} that the peculiar velocity of the IGM near the virial radius of galaxies is primarily determined by the gravitational potential of the halo. These optimizations or approximations could further reduce the computational cost by a factor of a few, but would not alter the fundamental scaling of the total computational cost. 

% However, the above optimization would only reduce the computational cost by a constant factor of a few and not change the fundamental scaling of the computation, $\mathcal{O}(\mathcal{N}_{\rm grid})$. 

\section{Validation in a Cosmological Simulation} \label{sec:test}

\subsection{Nyx simulation}

We further apply our analytic model to more realistic IGM data to evaluate its performance in practical settings and to gain initial insights into Ly$\alpha$ intensity maps. For this, we employ Nyx, a massively parallel cosmological $N$-body/hydrodynamics code introduced by \citet{2013ApJ...765...39A}. Nyx is particularly well-suited for modeling the intergalactic medium, as it omits particle-particle gravity calculations—an expensive but unnecessary step for simulating the diffuse IGM.

Using Nyx, we simulate a cubic volume of $l_{\rm box} = 20~h^{-1}~{\rm Mpc}$ with periodic boundary conditions on a $1024^3$ grid, down to redshift $z = 3$. This setup resolves structures with $20~h^{-1}~{\rm kpc}$ cell resolution, ensuring convergence in Ly$\alpha$ opacity statistics in the IGM \citep{2015MNRAS.446.3697L}. For the ionization state of the IGM, Nyx assumes a spatially uniform background from \citet{2012ApJ...746..125H}. As shown in \citet{2015MNRAS.446.3697L}, the choice of ultraviolet background (UVB) has little impact on Ly$\alpha$ opacity at $z\sim3$ as measured by the Ly$\alpha$ forest. 

\subsection{Mock IGM}

\begin{figure}
  \begin{center}
    \includegraphics[scale=0.6]{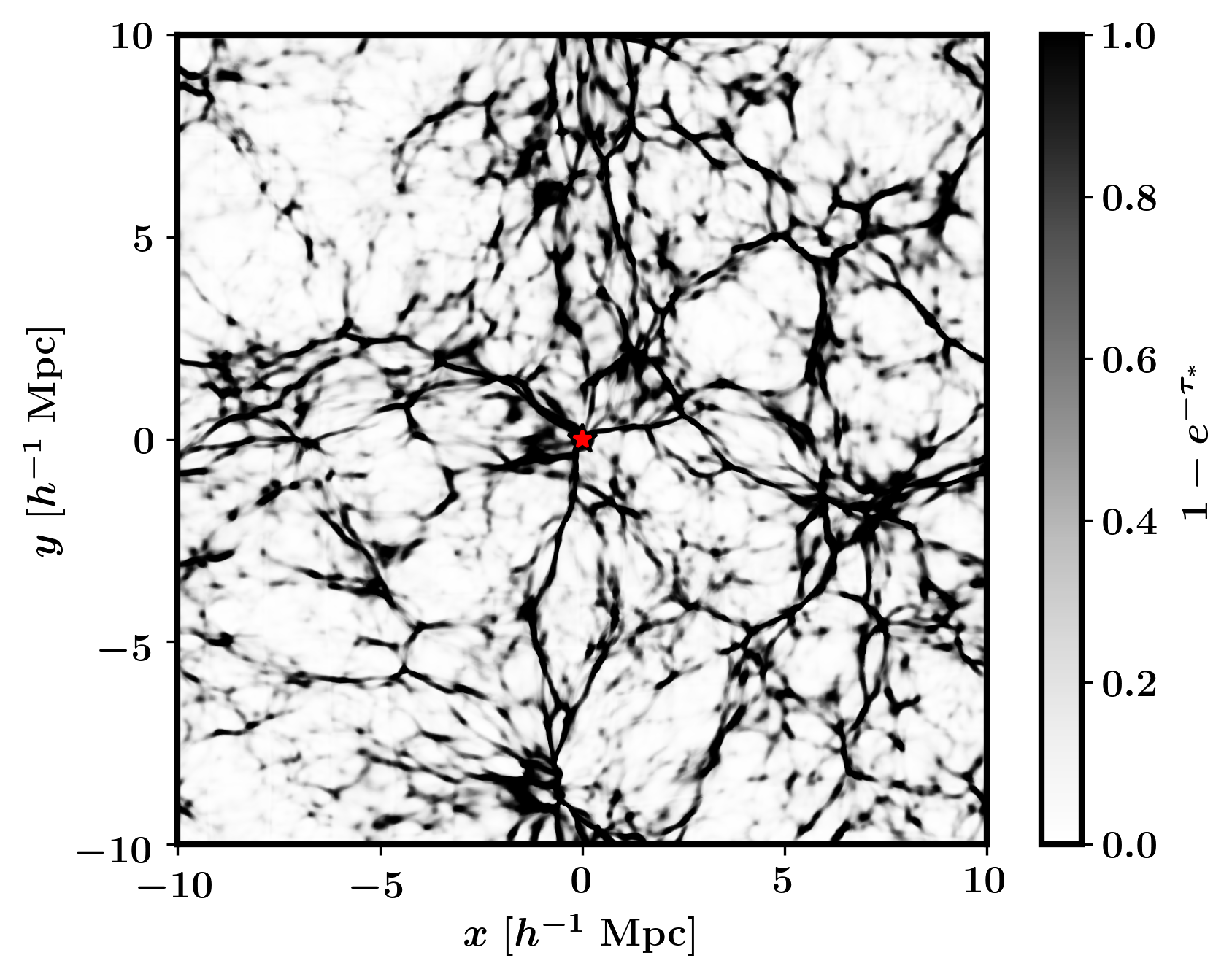}
  \caption{Scattering fraction at the Ly$\alpha$ resonance, $1-e^{-\tau_*}$, on a slice of IGM through the center of Nyx simulation at $z=3$. The scattering fraction decreases from 1 to 0 as the color transitions from black to white. The ionizing source at the center is marked with a star symbol.}
   \label{fig:Sr}
  \end{center}
\end{figure} 

We use the $z = 3$ snapshot of the Nyx simulation to calculate Ly$\alpha$ maps in post-processing, using the baryonic density and velocity fields. For this test calculation, we place a source at the most massive halo ($10^{12}~M_\odot$) identified within the box. For display purposes, we visualize the results with the source halo at the center, applying periodic boundary conditions, and evaluate observables in the reference frame of the source galaxy.

We smooth all the data originally on the $1024^3$ mesh onto a $256^3$ mesh, yielding a cell size around $0.1~{\rm Mpc}$, which is sufficiently small given that the photons random walk by about $0.1~{\rm Mpc}$ during the scattering process, as we observed in Figure~\ref{fig:trajectories}. Using Equation~(\ref{eq:tau_star}), we convert $n_{\rm HI}$ into Ly$\alpha$ transmission rate across the resonance, $e^{-\tau_*}$, and obtain the volume-averaged transmission rate of $\left< e^{-\tau_*} \right>_V\approx0.7$, which is consistent with measurements from the Ly$\alpha$ forest \citep{2013MNRAS.430.2067B}. The complement of the transmission rate, approximately 0.3 in this case, contributes to the scattered Ly$\alpha$ intensity maps. 

Figure~\ref{fig:Sr} shows the scattering fraction on the $xy$-plane passing through the center of the simulated IGM. While most of the volume is transparent to Ly$\alpha$ photons, the opacity along filaments is high enough to scatter nearly 100\% of the photons. At this epoch ($z=3$), the mean cosmic HI density is approximately $0.9 \times 10^{-10}~{\rm cm}^{-3}$, resulting in $\tau_* \approx 1.2$ (Eq.~\ref{eq:tau_star}). Since the UVB is uniform in the Nyx simulation, $n_{\rm HI}$ scales with the square of the density, allowing even modest overdensities in filamentary structures to raise $\tau_*$ well above unity.\footnote{In the real Universe, the ionizing background is likely higher in overdense environments to some degree due to the greater concentration of galaxies, making $\tau_*$ lower than in our uniform ionizing background case. It would be an important follow-up to investigate how a spatially varying ionizing background is reflected in the Ly$\alpha$ intensity maps.}

\subsection{Model vs. MCRT simulation comparison} \label{sec:comparison}

\begin{figure*}
  \begin{center}
    \includegraphics[scale=0.7]{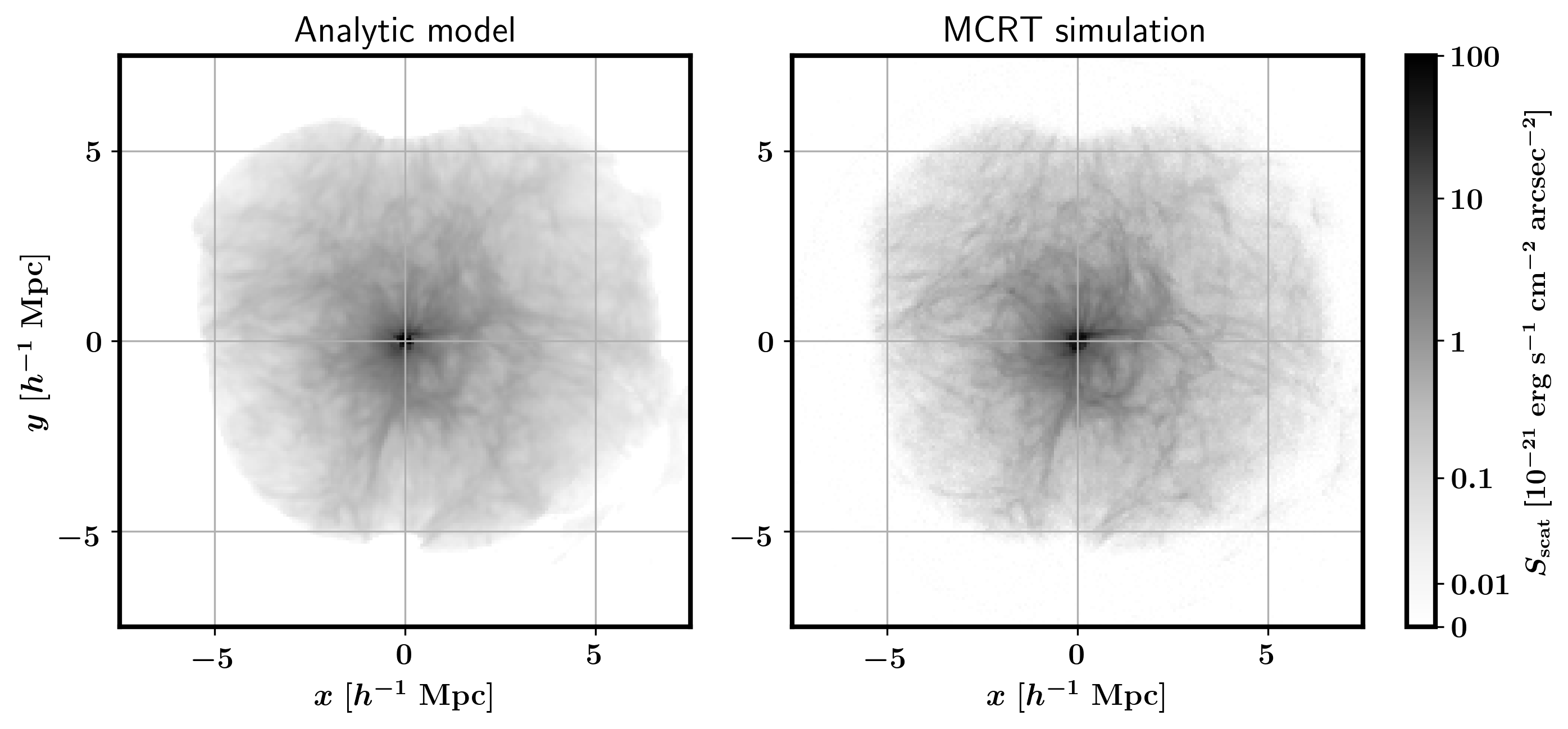}
  \caption{Comparison of surface brightness maps constructed from the analytic model (left) versus the MCRT simulation (right) from Sec.~\ref{sec:test}. The maps are generated by projecting the Ly$\alpha$ intensity distribution along the $z$-axis, with the $xy$ plane representing the sky.}
   \label{fig:LyaImapCompare}
  \end{center}
\end{figure*} 

\begin{figure}
  \begin{center}
    \includegraphics[scale=0.7]{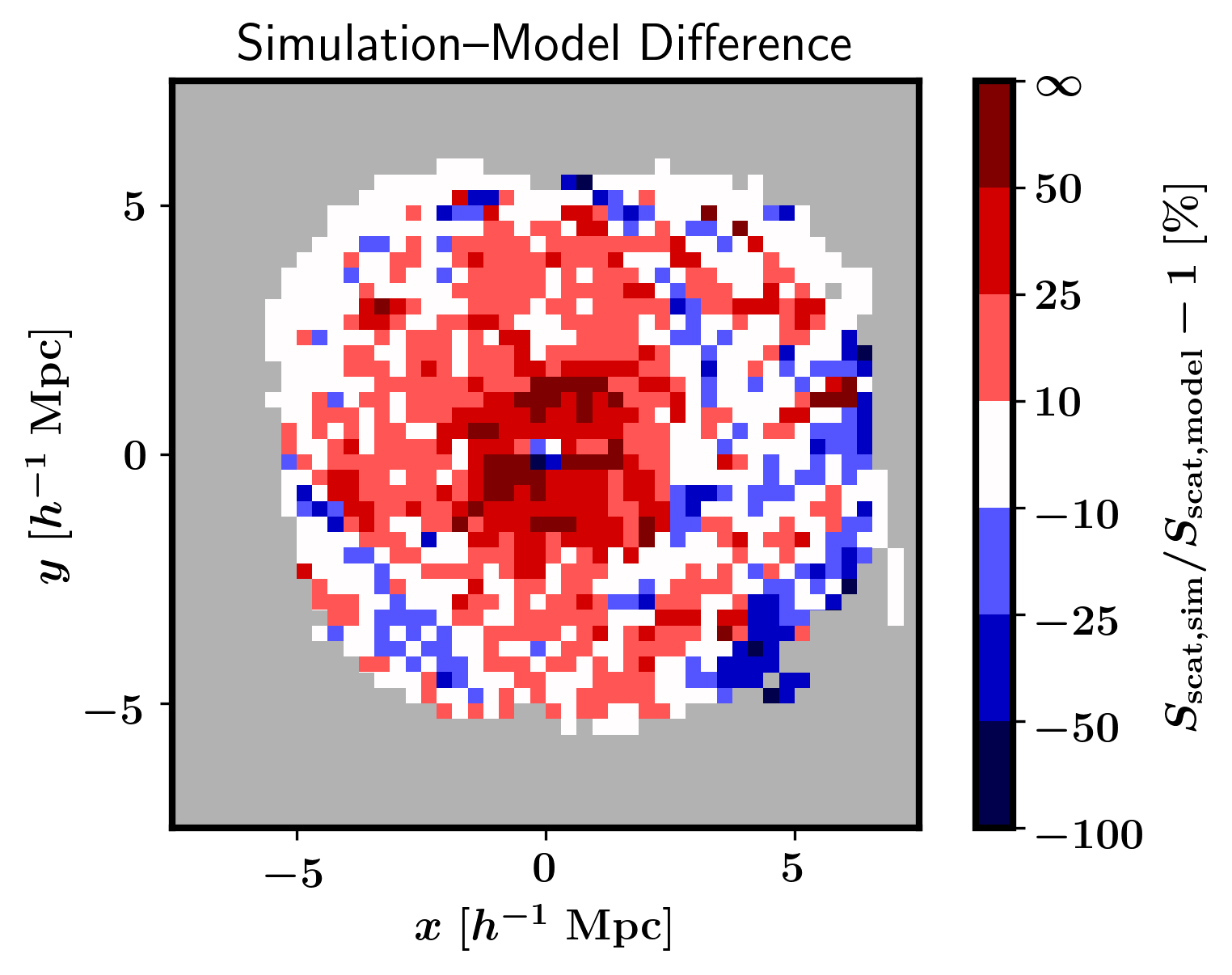}
  \caption{Fractional difference in Ly$\alpha$ intensity maps between the simulation and analytic models shown in Fig.~\ref{fig:LyaImapCompare}. To suppress numerical noise, the original $256^3$ maps are smoothed to $64^3$ resolution before taking the ratio, using the analytic model values as the denominator, giving a cell size of around $0.3~h^{-1}~{\rm Mpc}$. Red and blue indicate regions where the analytic model underestimates and overestimates the MCRT simulation results, respectively. We masked the cells with negligible intensity ($<10^{-23}~{\rm erg~s^{-1}~cm^{-2}~arcsec^{-2}}$) in gray.}
   \label{fig:LyaImapDiff}
  \end{center}
\end{figure} 

Figure~\ref{fig:LyaImapCompare} shows the resulting surface brightness ($S_{\rm Ly\alpha}$) of the scattered radiation, as estimated from our analytic model (Eq.~\ref{eq:S}) and from our MCRT simulation using $\approx1.5\times 10^7$ photons. Our analytic calculation takes less than a minute on typical personal computers while the MCRT simulation takes $\sim 10$ CPU-hours on typical computing clusters. 

We assume a flat luminosity per wavelength of $L_\lambda=10^{46}~{\rm erg}~{\rm s}^{-1}/\lambda_\alpha$ within the range $-600 < v_{\alpha,{\rm i}} < 300~{\rm km}~{\rm s}^{-1}$, and $L_\lambda=0$ outside the that range, for Ly$\alpha$ photons emitted at the center of the simulation box. Notably, photons emitted redward of resonance in the source’s rest frame can still encounter Ly$\alpha$ resonance in the IGM due to gravitational infall motions, up to the circular velocity of the halo ($\sim300~{\rm km}~{\rm s}^{-1}$ for a $10^{12}~M_\odot$ halo; see, e.g., Sec.~3.1 of \cite{2021ApJ...922..263P}). In the MCRT simulation, $v_{\alpha,{\rm i}}$ is drawn randomly with uniform probability within the same velocity range, with photons generated at the virial radius of the halo and initially propagating radially outward. For projection onto the sky plane (Eq.~\ref{eq:S}), we integrate along the $z$-axis over the full extent of the box (i.e., $r_{\parallel,{\rm l}} = -L_{\rm box}/2$ and $r_{\parallel,{\rm u}} = L_{\rm box}/2$).

We generally find good agreement between the model and the simulation. Due to much higher scattering rate in the filamentary structures (see Fig.~\ref{fig:Sr}), the cosmic web around the source is imprinted in the projected intensity maps. The sharp cutoffs in intensity are caused by the imposed cutoff in source intensity at $v_{\alpha,{\rm i}}=-600~{\rm km}~{\rm s}^{-1}$, and thus correspond to the last-scattering surface of the cutoff wavelength. These feature appear near identical in both maps. Aside from discreteness effects due to the finite number of photons in the simulation, there is no clear visual difference between the two results. We find that $\sim 37~\%$ of the total radiation from the source is converted into scattered radiation in both calculations.

\begin{figure*}
  \begin{center}
    \includegraphics[scale=0.7]{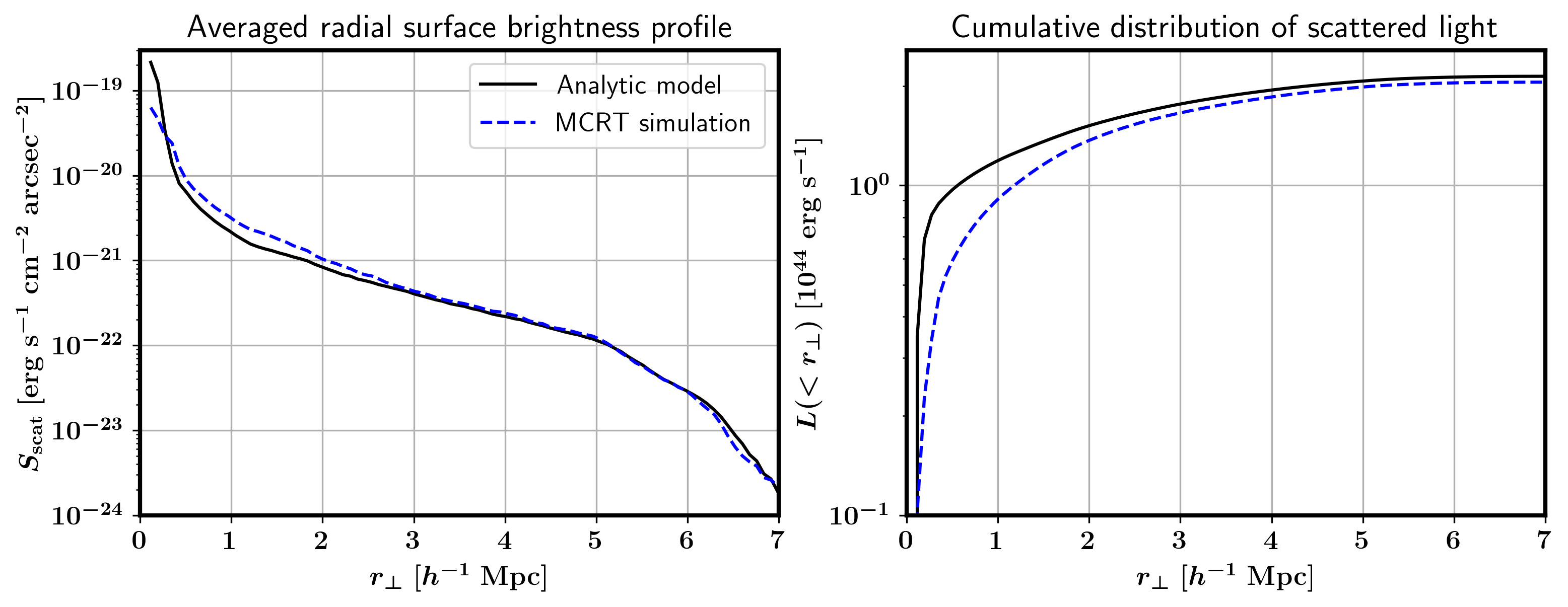}
  \caption{Comparison of the angle-averaged radial surface brightness profile of the Ly$\alpha$ intensity map from Fig.~\ref{fig:LyaImapCompare}. The result from the analytic model is shown as a black solid line, while the result from the MCRT simulation is shown as a blue dashed line.}
   \label{fig:Ir}
  \end{center}
\end{figure*}

The fractional difference in intensity (Fig.~\ref{fig:LyaImapDiff}) and the comparison of the radially averaged surface brightness profiles and cumulative energy distributions in Figure~\ref{fig:Ir} reveal subtle discrepancies between the analytic model and the simulation. The radial surface brightness profile (left panel of Fig.~\ref{fig:Ir}) shows that the analytic model overestimates the intensity by up to a factor of three at $r_\perp \lesssim 0.3~h^{-1}~{\rm Mpc}$, while underestimating it by up to a factor of 1.5 at $0.3 \lesssim r_\perp \lesssim 3~h^{-1}~{\rm Mpc}$. In Figure~\ref{fig:LyaImapDiff}, this trend appears as a few blue pixels at the center\footnote{Note that the cell size in the figure is also around $0.3~h^{-1}~{\rm Mpc}$.} surrounded by red or dark red regions.

The cumulative distribution (right panel of Fig.~\ref{fig:Ir}) shows that the discrepancies in these two ranges largely cancel out within $r_\perp \approx 3~h^{-1}~{\rm Mpc}$, resulting in less than a 10\% error in the total scattered radiation at $r_\perp \gtrsim 3~h^{-1}~{\rm Mpc}$. The analytic model predicts a slightly more compact and brighter core near the center, while still conserving the total number of photons to within 10\%. The fractional differences in individual cells are generally larger but remain within 30\% at $r_\perp \gtrsim 3~h^{-1}~{\rm Mpc}$ (see Fig.~\ref{fig:LyaImapDiff}).

\begin{figure}
  \begin{center}
    \includegraphics[scale=0.41]{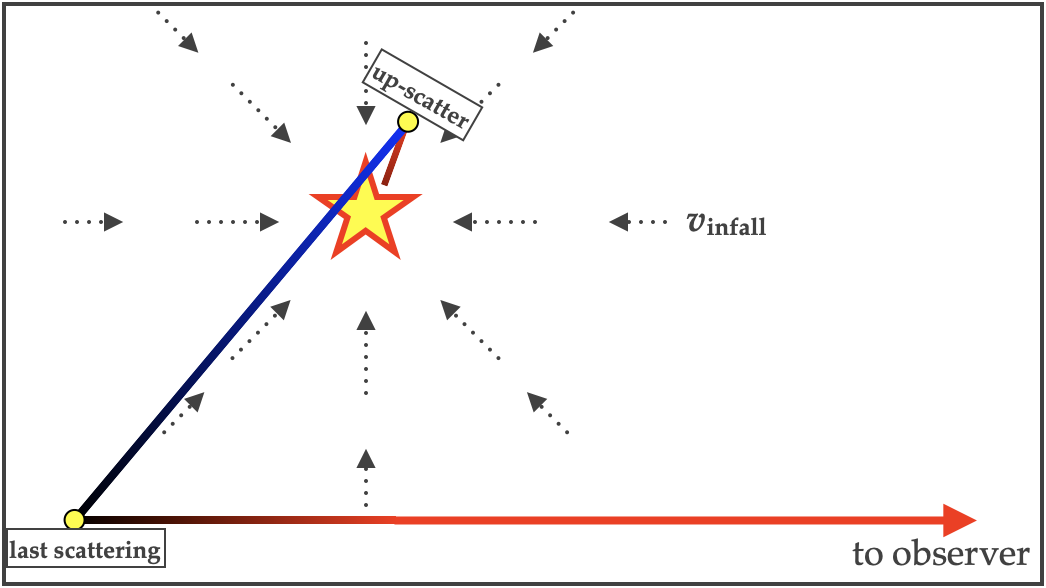}
  \caption{Schematic illustration of the Ly$\alpha$ scattering process near the source, where strong gravitational infall motion of the IGM can significantly blue-shift photons to the blue side of the Ly$\alpha$ resonance. The star symbol marks the source location. The solid arrow indicates a hypothetical path of a typical Ly$\alpha$ photon scattered in the vicinity of the source galaxy. The color gradient along the photon path denotes changes in the photon wavelength due to cosmological redshift and scattering. The dotted arrows represent the infall motion of the IGM surrounding the source.}
   \label{fig:upScattering}
  \end{center}
\end{figure}

We attribute the difference at $r_\perp \lesssim 3~h^{-1}~{\rm Mpc}$ to the strong peculiar infall velocity of the IGM, redistributing the photons over a larger volume. Photons emitted near or slightly redward of resonance are typically scattered by the strongly infalling IGM in the vicinity of halos. During the scatter, they can be substantially blueshifted in the source's frame if the scattering direction aligns with the infall motion. These photons can then travel some distance, redshifting back toward resonance, and be scattered again. Given that the infall velocity can reach up to $300~{\rm km}~{\rm s}^{-1}$, scattered photons can attain $v_\alpha = -300~{\rm km}~{\rm s}^{-1}$, thereby redistributing the location of their last scattering over a circular region with a radius of $3~h^{-1}~{\rm Mpc}$. This explains why the MCRT simulation produces lower surface brightness within the virial radius ($\sim0.3~h^{-1}~{\rm Mpc}$) of the halo and higher surface brightness at $0.3\lesssim r\lesssim3~h^{-1}~{\rm Mpc}$. This is a complex radiative transfer effect due to strong peculiar velocity gradients around the virial radius, unaccounted for in our analytic model, and not present in the static IGM case discussed earlier. A schematic illustration of this process is shown in Figure~\ref{fig:upScattering}. 

Overall, the performance of our analytic model appears satisfactory on IGM scales. The discrepancy caused by peculiar velocity effects is confined to regions close to the source and does not significantly modify the total amount of scattered radiation. In practice, these regions are likely to be under-resolved or masked together with the source galaxy in most surface brightness surveys. Moreover, the discrepancy is expected to be even smaller for fainter galaxies, which have weaker peculiar infall velocities. Therefore, we expect our analytic model to produce highly realistic Ly$\alpha$ intensity maps in typical environments.

\section{Discussion and Conclusions} \label{sec:summary}

{\it Summary:} We have developed an analytic model to generate a Ly$\alpha$ intensity map from mock IGM and Ly$\alpha$ sources. We validated the model using a simulated IGM and found good agreement with MCRT simulations. Our approach provides a complementary tool for generating Ly$\alpha$ intensity maps from cosmological simulations, enabling efficient exploration of the physical processes shaping Ly$\alpha$ emission. In future work, we plan to apply this model to generate mock Ly$\alpha$ intensity maps on cosmological scales, targeting ongoing Ly$\alpha$ mapping surveys such as MIRACLES. 

{\it Applications:} Our analytic model enables fast and straightforward experimentation with different assumptions about the spectral shape around Ly$\alpha$ for individual galaxies and quasars, as well as the spatial variation of the ionization rate. The relation between the size of the last-scattering surface and the initial wavelength offset of Ly$\alpha$ photons at emission, shown in Figure~\ref{fig:LSS}, highlights the critical role of the spectral shape blueward of Ly$\alpha$ in shaping the large-scale intensity distribution. Introducing strong ionizing sources, such as quasars, can raise the local ionizing background and significantly reduce the scattered fraction of Ly$\alpha$ photons in proximity zones and surrounding regions. 

Moreover, the scattered radiation is expected to correlate strongly with the large-scale structure of the Universe beyond the size of the scattering surface ($\gtrsim 10~{\rm Mpc}$). Therefore, another key objective is to quantify the relationship between the underlying matter density and the scattering emissivity, analogous to the theoretical framework developed by \citet{2014ApJ...786..111P} for other Ly$\alpha$ intensity components. Such a relationship may even enable the reconstruction of the intensity map from the dark matter distribution through a sequence of straightforward operations such as Fourier transformation and convolution. We plan to explore these directions in future work.

{\it Caveats:} It is important to note that our model is applicable only to the post-reionization Universe, where damping-wing opacity is negligible over most of the volume, and to the low-density IGM, where peculiar velocities are weak enough that Doppler shifts of scattered photons remain small. In denser environments such as the circumgalactic medium (CGM) or interstellar medium (ISM), the assumptions underlying our model no longer hold, and realistic Ly$\alpha$ intensity maps must be generated through full numerical simulations. Once these photons enter the IGM, where the density is lower and peculiar velocities are weaker, our analytic prescription can then provide an accurate prediction for the scattered intensity from these photons.

In this work, we tested our model for the contribution from a single source in a massive halo ($10^{12}~M_\odot$), which showed a mild deviation of the analytic result from the simulation within $\lesssim 1~{\rm Mpc}$ of the halo, due to strong peculiar velocity gradients. While most of the cosmic volume exhibits much weaker velocity gradients, many upcoming surface brightness surveys would likely target overdense regions to detect stronger signals. A potential strategy to address this challenge is to adopt a hybrid approach: using MCRT to simulate photons that are expected to undergo complex scattering—such as those emitted near resonance (e.g., $v_{\alpha,{\rm i}} \gtrsim -100~{\rm km}~{\rm s}^{-1}$) from massive halos (e.g., $\gtrsim 10^{12}~M_\odot$)—while applying the analytic prescription to other photons to save computational cost without sacrificing accuracy. Future studies will be needed to assess how well the analytic model reproduces MCRT results for numerous galaxies in such dense environments and to determine the optimal strategy for overcoming these limitations effectively.

Additionally, we note that our analysis focused on radiation emitted near resonance ($v_{\alpha,{\rm i}} \gtrsim -1000~{\rm km}~{\rm s}^{-1}$), but far-bluer photons ($v_{\alpha,{\rm i}} \ll -1000~{\rm km}~{\rm s}^{-1}$) from the galactic continuum can also contribute to scattered radiation at more distant locations. For such photons, a proper treatment must account for the redshift evolution of $H(z)$ and potential scattering when they shift to other Lyman-series lines. While we do not derive this contribution here, it can add an extra component to the intensity field that correlates with the IGM density through the opacity term, $1 - e^{-\tau_*}$, but not with the local sources.

{\it Comparison to previous works:} For the fully neutral IGM with significant damping-wing opacity, analytic solutions have been presented by \citet{1999ApJ...524..527L} and \citet{2025arXiv250101928S}, which are more appropriate for epochs at $z\gtrsim10$, when the Universe is expected to remain largely neutral. In such environments, the damping wing introduces substantial opacity over a broad wavelength range beyond the resonance core, which is broadened primarily by thermal motion. Consequently, Ly$\alpha$ photons tend to scatter over larger physical distances.\footnote{Notably, our solution for monochromatic sources (Fig.~\ref{fig:rPjt}) closely resembles the free-streaming, large-$\nu$ cases in Fig. 5 of \citet{1999ApJ...524..527L}. This similarity arises because their large-offset cases effectively probe photons that last scattered far redward of resonance, forming a thin spherical shell—an effect analogous to the last-scattering surface in our monochromatic source calculations.} This work explores an entirely different regime, where the damping-wing opacity is negligible and scattering only occurs near the resonance.

{\it Conclusion:} This work reaffirms that, under certain astrophysical conditions, the complex process of Ly$\alpha$ resonant scattering can be effectively captured with simple analytic models. Our model offer valuable complements to numerical simulations, serve as benchmarks for testing radiative transfer codes, and provide pedagogical insights for interpreting current and future Ly$\alpha$ observations. Future work will include multiple sources, spatially-varying UV backgrounds, and explicit modeling of survey selection effects.

~~

\section*{Acknowledgment}
Authors thank the referee, Z. Zheng, for his constructive comments that significantly improved this manuscript. HP thanks K. Nagamine for his helpful comments. HP was supported by the World Premier International Research Center Initiative (WPI), MEXT, Japan and JSPS KAKENHI Grant Number 19K23455. HP was supported in part by NSF grant PHY-2309135 to the Kavli Institute for Theoretical Physics (KITP). Numerical simulations for this work were performed on the idark computing cluster of the Kavli Institute for Physics and Mathematics of the Universe, the University of Tokyo. CB was supported by the JSPS as International Research Fellow during this project. HY was supported by MEXT/JSPS KAKENHI grant number 21H04489, JST FOREST Program grant number JP-MJFR202Z, and Strategic Professional Development Program for Young Researchers, TRiSTAR fellow.

\section*{Data Availability}
The Ly$\alpha$ MCRT code used in this work is available at \url{https://github.com/hcosmosb/LyaMC}. 
The Nyx simulation code is available at \url{https://github.com/amrex-astro/Nyx}. 
Other simulation output data used in this study are available from the first author upon reasonable request.

\bibliographystyle{apj}
\bibliography{reference}

\appendix

% #($\sigma_\alpha$), which is a function of the wavelength difference w.r.t. the Ly$\alpha$ resonance ($\lambda_\alpha$) and the IGM temperature ($T$)
\section{Ly$\boldsymbol{\alpha}$ Opacity of a Sightline in a Cosmologically Expanding Space}

In this section, we provide the full derivation of Equation~(\ref{eq:tau_star}). The Ly$\alpha$ opacity is given by integrating the product of the HI number density and the Ly$\alpha$ cross-section:
\bea
\tau_* = \int^{\infty}_{0} n_{\rm HI}(\mathbf{r}) \sigma_\alpha(\Delta\lambda_\alpha,T) a(z)\text{d}r \, .
\eea
Here, $\sigma_\alpha$ is the Ly$\alpha$ cross-section, expressed as a function of the wavelength offset from the Ly$\alpha$ resonance, $\Delta\lambda_\alpha \equiv \lambda - \lambda_\alpha$, and the IGM temperature, $T$. The comoving position along the sightline is given by $\mathbf{r} = r\mathbf{\hat{r}}$, where $\mathbf{\hat{r}}$ is the unit vector along the propagation direction. The wavelength offset can be rewritten as
\bea \label{eq:dla}
\Delta \lambda_\alpha (\mathbf{r}) = \Delta \lambda_{\alpha,{\rm i}} + \lambda_\alpha \left[ \frac{ra(z)H(z)+ v_{{\rm pe},r}(\mathbf{r})}{c}\right],
\eea
where $\Delta \lambda_{\alpha,{\rm i}}$ is the initial wavelength offset, $H(z)$ is the Hubble expansion rate, and $v_{{\rm pe},r}(\mathbf{r}) \equiv \mathbf{\hat{r}} \cdot \mathbf{v}_{\rm pe}(\mathbf{r})$ is the line-of-sight component of the IGM peculiar velocity at location $\mathbf{r}$ relative to the emission source, accounting for the Doppler shift of photons in the IGM frame. 

The functional form of $\sigma_\alpha$ follows the Voigt function, which arises from the convolution of the Lorentzian damping-wing profile of individual atoms and the Gaussian resonance core profile due to thermal motion, giving
\bea
\sigma_\alpha(\Delta\lambda_\alpha,T) = [5.889\times 10^{-14}~{\rm cm}^2] \left[\frac{T}{10^4~{\rm K}}\right]^{-0.5} \phi(x).
\eea
Here, 
\bea
\phi(x)\equiv \frac{a_\nu}{\pi} \int^{\infty}_{-\infty} \frac{e^{-y^2} \text{d}y}{[y-x]^2+a_\nu^2} \approx
\begin{cases}
e^{-x^2} & (|x| \lesssim 3;~\text{resonance core}) \\
\dfrac{a_\nu}{\sqrt{\pi} x^2} & (|x| \gtrsim 3;~\text{damping wing})
\end{cases}
,
\eea
is the Voigt profile, where $a_\nu \equiv 4.7\times10^{-4}[T/10^4~{\rm K}]^{-1/2}$ is the Voigt parameters, $x(\mathbf{r}) \equiv -[\Delta\lambda_\alpha/\lambda_\alpha]/[v_{\rm th}/c]$ is the dimensionless wavelength offset and $v_{\rm th}$ is the thermal velocity, given in Equation~(\ref{eq:v_th}). 

In this study, we consider the case in which a photon is initially emitted blueward of the Ly$\alpha$ resonance, beyond the thermal broadening scale ($x \gg 3$), and redshifts through the resonance core ($-3 \lesssim x \lesssim 3$) to the red side, eventually reaching wavelengths well beyond the thermal broadening scale ($x \ll -3$). At IGM scales ($\gtrsim {\rm Mpc}$), the width of the Gaussian core is negligibly small, allowing us to use the Sobolev approximation for the line profile $\phi(x)$ and treat it as a Dirac delta function: $\phi(x) \approx \sqrt{\pi}\,\delta^{\rm D}(x)$.\footnote{At galactic and smaller scales, deviation from this assumption becomes significant \citep{2022MNRAS.517....1S}.} This Dirac Delta function is evaluated at the scattering location $\mathbf{r} = \mathbf{r}_{\rm s}$, yielding 
\bea
\tau_* = \sqrt{\pi}~n_{\rm HI} (\mathbf{r}_{\rm s}) a(z) \left[\frac{T}{10^4~{\rm K}}\right]^{-0.5} \left| \frac{\text{d}x}{\text{d}r}(\mathbf{r}_{\rm s}) \right|^{-1} [5.889\times 10^{-14}~{\rm cm}^2].
\eea
Inserting Equation~(\ref{eq:dla}) into the term $\text{d}x/\text{d}r$ in the above expression gives:
\bea
\tau_* = \left[\frac{n_{\rm HI}(\mathbf{r}_{\rm s})}{0.75\times10^{-10}~{\rm cm}^{-3}}\right] \left[\frac{H(z)+[1+z]\text{d}v_{{\rm pe},r}/\text{d}r}{301~{\rm km}~{\rm s}^{-1}~{\rm pMpc}^{-1}}\right]^{-1},
\eea
where pMpc stands for physical megaparsec. Assuming a uniform ($n_{\rm HI}(\mathbf{r}_{\rm s}) = n_{\rm HI}$) and static IGM ($v_{\rm pe} = 0$), and neglecting the dark energy term in the Hubble rate (Eq.~\ref{eq:H}; valid for $[1+z]^3 \gg 1$), simplifies the expression to:
\bea
\tau_* \approx \left[\frac{n_{\rm HI}}{0.75\times10^{-10}~{\rm cm}^{-3}}\right] 
\left[\frac{h}{0.678}\right]^{-1} \left[\frac{\Omega_{\rm M}}{0.307}\right]^{-0.5} 
\left[\frac{1+z}{4} \right]^{-1.5}.
\eea
This is Equation~(\ref{eq:tau_star}).

% where $\mathbf{r}_{\rm s} = r_{\rm s}\,\mathbf{\hat{r}}$ denotes the scattering location

\end{document}